\documentclass[a4paper,12pt]{extarticle}
\usepackage{ed}
\usepackage{url}
\usepackage{hyperref}
\nocolon

\graphicspath{{./pic}}

\def\twopicts#1#2{%
\centerline{\hbox to \columnwidth{\hss\includegraphics[width=0.5\textwidth]{#1}\;%
    \includegraphics[width=0.5\textwidth]{#2}\hss}}}

\begin{document}
\selectlanguage{english}
\title{Some problems of developing astrophysical equipment and combining it with optical
telescopes}
\author{Emelianov E.V.}
\date{\small\it Special Astrophysical Observatory of the Russian Academy of Sciences, Nizhny
Arkhyz, 369167, Russia}

\maketitle
\abstract{
The results of a study of the accuracy characteristics and image quality on the SAO RAS optical
telescopes, Zeiss-1000 and BTA, using the recently developed ``Telescope Analyzer'' device are
described: a method for determining the coefficients of the pointing correction system, the
position of the aberration axis along the coma in images of star fields, natural frequencies of
vibrations of mechanical systems, prospects for the development of the device. Attention is paid
to the thermal deformations of the BTA main mirror and measures to reduce them. Mention
is made of systems being developed for partial correction of wavefront aberrations
due to imperfect mechanics, and plans to modernize the control system of the BTA. The complex of
0.5-meter telescopes ``Astro-M'' has not been forgotten: hardware and software solutions for
automating the first and second telescopes, plans for commissioning of telescopes No~3--5 are
described. Links to repositories with developed software and hardware products are provided.
    %\doi{10.26119/VAK2024-ZZZZ}
}

\section{Towards the problem of scientific experiment automation}
When developing various astrophysical equipment, one has to deal with both the lack of
ready-made blocks and the lack of GNU/Linux SDK (software development kit) or even a clear
documentation with description of the operating protocol of purchased devices. As an example of
the second, the HSFW turrets (Edmund Optics) control tool developed by reverse
engineering\footnote{\url{https://github.com/eddyem/eddys_snippets/tree/master/HSFW_management}},
another tool\footnote{\url{https://github.com/eddyem/CCD_Capture}} was developed to unify the
acquisition of images from various CCDs and CMOS (since no supplier offered ready-made
command line utilities for working with their light receivers). Here we will mainly talk about hardware
developments.

\begin{pict}
    \twopicts{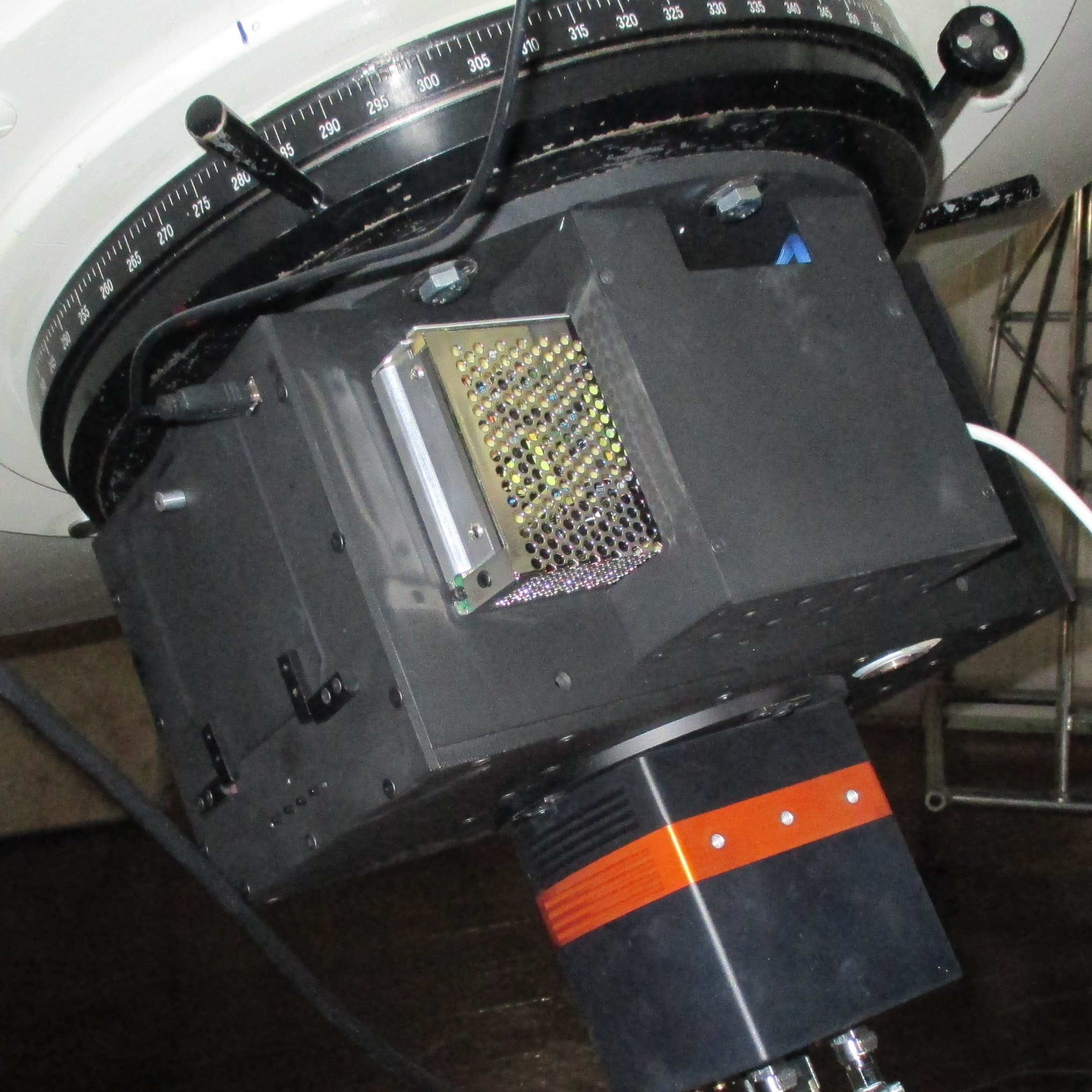}{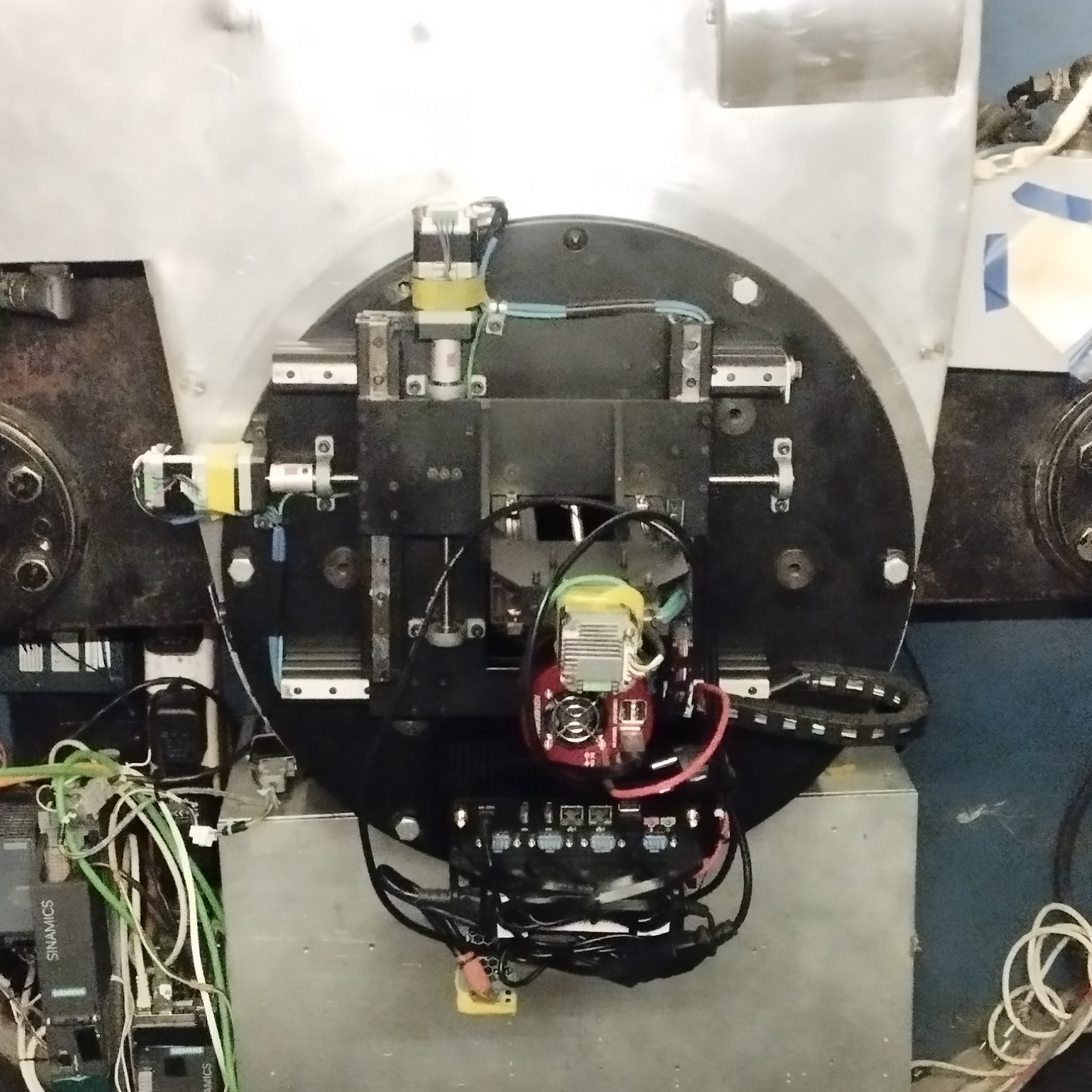}
    \caption{MMPP mounted on Cassegrain focus of Zeiss-1000 (left) and TeA on BTA primary
    focus (right).}
    \label{mmpptea}
\end{pict}

\subsection{Steppers. Steppers are everywhere!}
The more complex the equipment, the more moving parts it contains: rotating (turrets and rotating
stages), moving (actuators and linear stages) and their combinations (cam mechanisms, etc.).

For example, Multi-mode photometer-polarimeter (MMPP)~\cite{MMPP} for Zeiss-1000 telescope
contains two HSFW filter turrets and two linear stages with one rotating stage on each (for
wave-plate and polaroid). As stated above, HSFW management control required only reverse
engineering, but those moving stages required full development of a new device controlled two
steppers\footnote{\url{https://github.com/eddyem/mmpp}}. Commercially available at that time
stepper motor controllers were too large to fit inside the device housing. The developed PCBs were
small enough, controlled over UART through simple USB-UART converter. PCBs were designed
for using with DRV8825-based steppers controller boards (unfortunately, even at that time, a
significant proportion of developments was based on Aliexpress; nowadays it is close to 100\%).

\begin{figure}[t]\centering
    \includegraphics[width=0.7\textwidth]{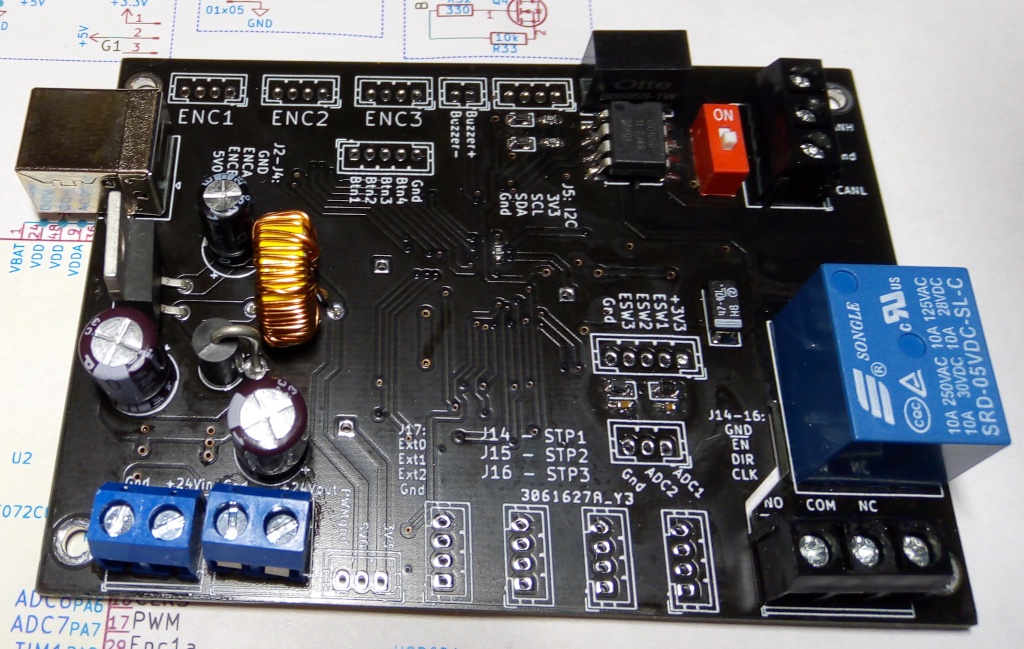}
    \caption{Controller of three stepper motors with feedback.}
    \label{3step}
\end{figure}

The next more complex device was the ``Telescope Analyzer'' (TeA), the basis of which is a
three-coordinate moving stage with CMOS detector ZWO ASI 1600MM~\cite{TeA}.
To improve the precision characteristics of the device an optical encoder is installed on the
shaft of each stepper motor, which allows you to control skipping steps. Thanks to this, the
positioning accuracy increased to 25\,$\mu$m over full observation night; the main problem in the
repeatability of results ultimately became bending of structures, leading to a shift in the image
centre position by a distance of up to 80\,$\mu$m when the platform was rotated to 180\degr.
Stepper motor controllers (now commercial drivers have become quite compact) are also installed
on their housing. This task requires the use of a microcontroller that has at least three timers
capable of operating in encoder mode, and at least three more with PWM outputs to generate
microstepping pulses for each driver. The newly developed device has the ability to be controlled
both via USB and
CAN\footnote{\url{https://github.com/eddyem/stm32samples/tree/master/F0:F030,F042,F072/3steppersLB}}
(fig.~\ref{3step}). As with all our devices, when connected via USB, a standard USB-to-serial
converter PL2303 is simulated (instead of the classic CDC-ACM interface it was necessary to do
emulation so that some Linux distributions would not try to perceive the device as a modem when
connected and block it with the corresponding daemon). Since it is possible to simultaneously
connect several devices via USB, and all of them will have the same VID\slash PID, marking by
interface descriptor is used (and {\tt udev} daemon creates a corresponding symlink in the {\tt /dev}
directory when connecting). This method is also convenient for distinguishing between interfaces
of one device (one STM32 can have up to seven USB interfaces).
For ease of use, the USB-protocol is text.
Each line ends with a newline, so the user can either develop his own application to work with the
device via USB or directly enter commands and read responses in the terminal or even partially
automate this using bash scripts. The CAN protocol has a variable message length; in general, all
8~bytes of data are used for data transmission: 2~bytes~-- the command code, 1~byte~-- its
parameter (for example, channel number), 1~byte~-- the error code (in the device response),
4~bytes of data.

It worth saying that the vast majority of our later developed devices have at
least CAN (as the world's most reliable and usable interface), USB (as the most popular~-- you can
connect even with a smartphone), or both on board.

During the development and modernization of some spectrograph units ready-made commercial
solutions were tested. Thus, ``Pusirobot''\footnote{\url{https://en.pusirobot.com/}} drivers are based
on the CanOpen protocol which allows to significantly simplify the process of integration into the
equipment control system. This company offers a fairly extensive range of various stepper motor
controllers (including integrated solutions). And in general, if the motors are far enough apart
from each other, it is a very convenient solution to connect them with a CAN bus (where other
devices can be connected; you just need to make sure that there is no conflict of identifiers). As
usual, there was no ready-made software for controlling motors from the command line, so several
utilities have been developed\footnote{\url{https://github.com/eddyem/pusirobot}} to work with
them, including a socket server that allows multiple applications to work simultaneously on a serial
port or CAN-bus and simplest application to control any ``Pusirobot'' motor. We used these motors
in the  pre-slit section of the BTA N2 focus spectrographs (NES and MSS) for moving heavy
platforms with optical elements and rotating turrets.

\begin{pict}
    \twopicts{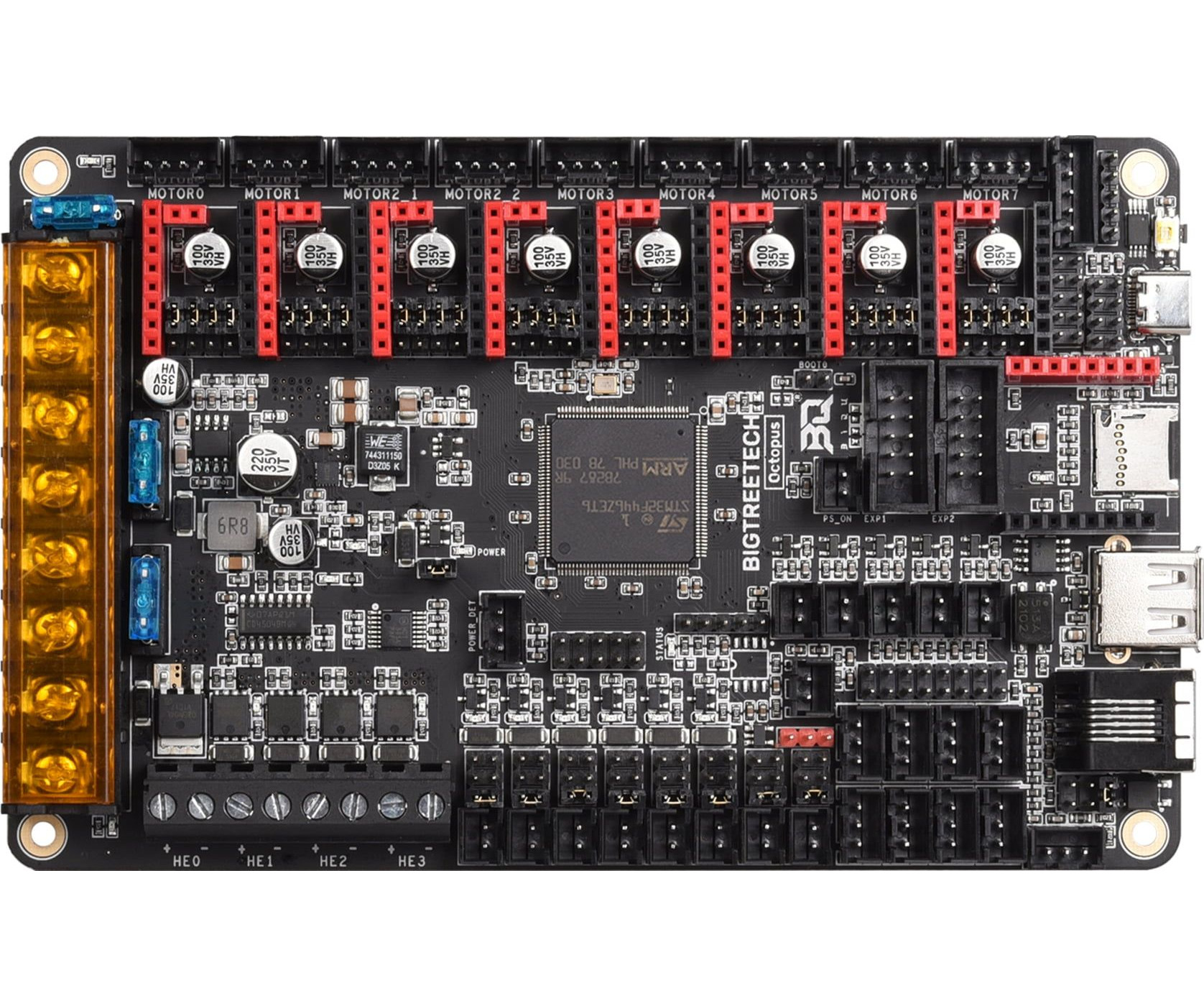}{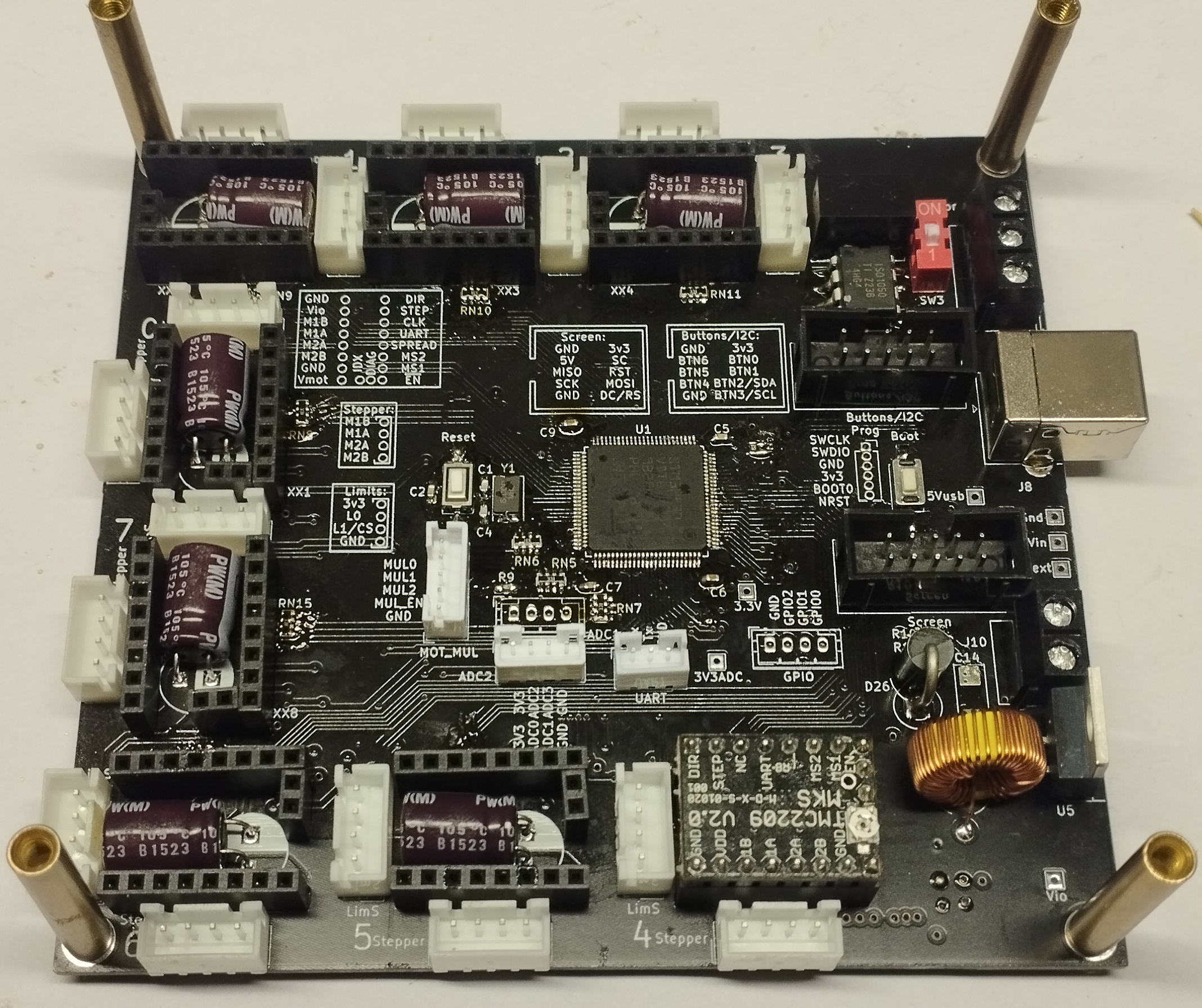}
    \caption{BigTreeTech ``Octopus'' (left) and our analogue (right).}
    \label{8step}
\end{pict}

Such ready-made solutions are convenient in terms of ease of implementation, but, unfortunately,
they have quite decent dimensions. Developing control system of the universal module of the
pre-slit part of spectrographs (mounted on telescope part of the fibre-fed spectrograph or the
whole spectrograph like ESPriF~\cite{ESPriF}), we had to deal with controlling a large number of
small-sized stepper motors. In common case up to eight steppers. With the advent of
mass-produced CNCs a proposal arose for controllers of such multi-axis machine tools. We
considered one of them, ``BigtTreeTech Octopus'', as the basis for controlling the pre-slit part.
Unfortunately, it turned out that this controller is not only unable to separately control eight stepper
motors, but also severely limits user to the standard protocol capabilities. Thus, another
development was born, a controller for eight stepper
motors\footnote{\url{https://github.com/eddyem/stm32samples/tree/master/F3:F303/Multistepper}}
(fig.~\ref{8step}). The controller allows to independently control the movement of all connected
motors. It has connectors for popular stepper motor drivers such as DRV8825, TMC2209,
TMC2130 and many other. Each motor can be serviced by two limit switches (or one, if an
SPI-controlled driver is used). SPI- or UART-controlled drivers allow user to programmatically
change the settings for current, microsteps and other parameters. However, despite the ``software''
ability to change the current, for low-power motors (100-300 mA) commercial driver modules must
be modified by changing of the sense resistors value.
USB protocol of this device use text commands, getter have only one word (which may ends with
number~-- getter parameter) and setter have integer parameter going after equal sign. CAN-bus
protocol based on variable length messages; the first two data bytes are function code, third is
parameter number, fourth~-- error code and the rest four bytes are setter\slash getter data. This
protocol turned out to be quite convenient and became established in many of our devices as a
universal one.

\begin{pict}
    \twopicts{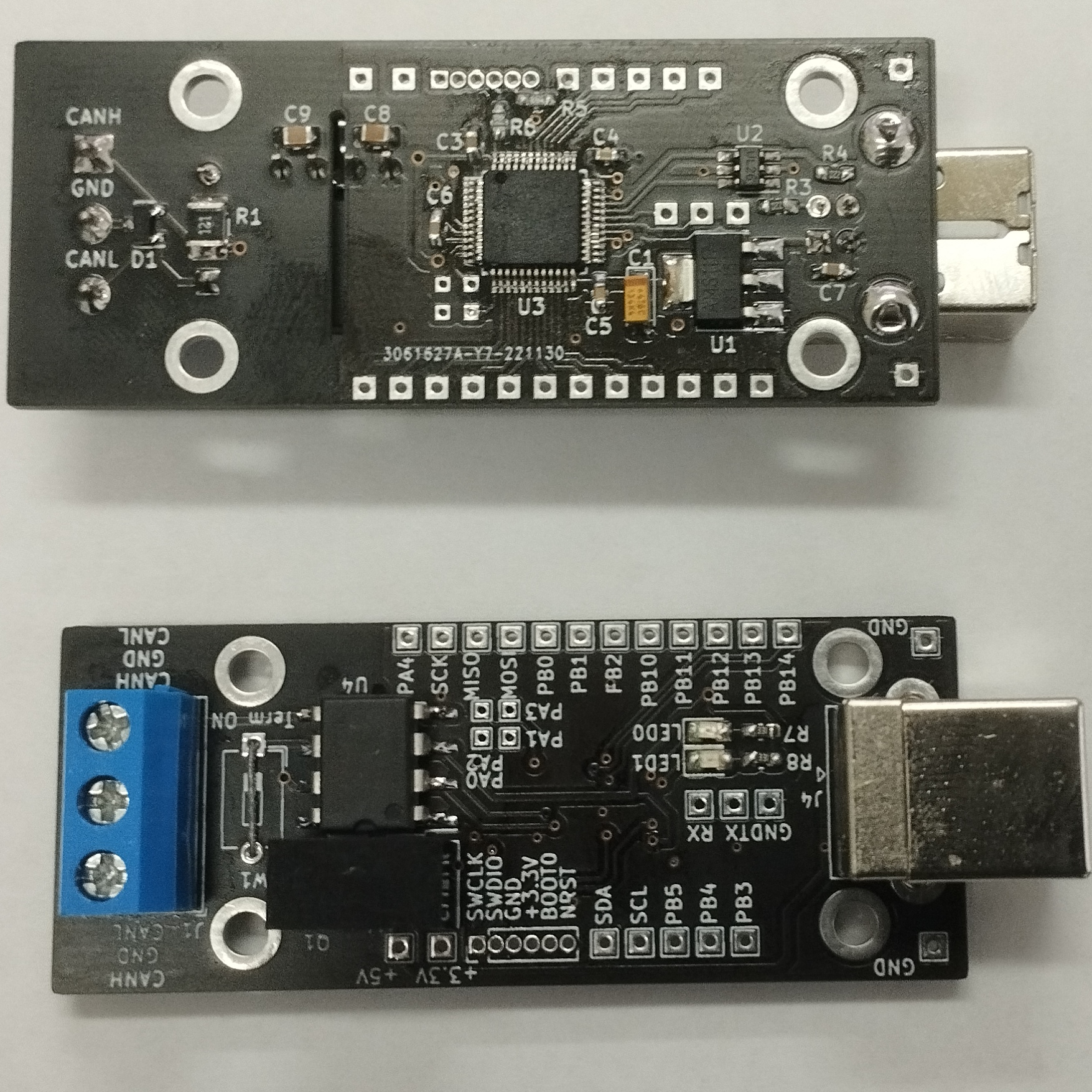}{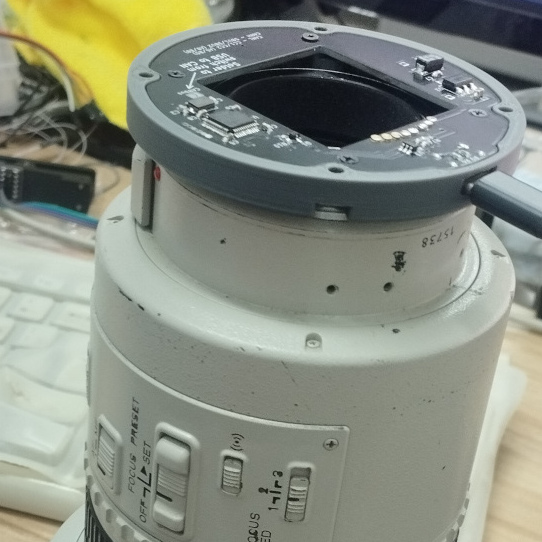}
    \caption{USB--CAN converter (left), Canon lens controller mounted on lens (right).}
    \label{canbuscanon}
\end{pict}

\subsection{And some other devices}
CAN has already been mentioned several times here but how to connect this bus to a regular
computer (the problem is especially critical for single-board and compact computers that do not
have PCI or PCI-x buses)? There are two options here: take a ready-made CAN-USB converter or
make your own. The second method is convenient because you can implement a completely
arbitrary  device protocol (and in practice, such devices are also useful because it is easy to
assemble a temporary solution in one evening on their basis, say, relay controller or
12-bit ADC over CAN-bus). And, oddly enough, it turned out that the galvanically isolated converter
has a cost lower than commercially offered analogues. STM32F072 was selected among the
inexpensive microcontrollers capable of performing this task. The
device\footnote{\url{https://github.com/eddyem/stm32samples/tree/master/F0:F030,F042,F072/usbcan_ringbuffer}}
 (fig.~\ref{canbuscanon}) was tested on speeds up to 250000 baud in full bus load mode showing
no stumble or packet loss. The device have simplest letter-based protocol over USB, allowing to
send CAN-packets of variable length with given ID (numerical data can be in decimal,
hexadecimal, octal or binary format). Also it allows to activate hardware CAN-filters and software
blacklist.

Now let's get back to commercial devices. When developing spectrographs with a small beam
diameter, it turned out that Canon lenses very well satisfy the task of imaging the spectrum of
an echelle-spectrograph on a light detector. The task of controlling these lenses required a fairly
long study of the data exchange protocol between camera and
lens\footnote{\url{https://github.com/eddyem/canon-lens}}. As a result of this work, a compact
device with a bayonet flange appeared, placed directly on the lens and allowing it to be controlled
either via USB or CAN
bus\footnote{\url{https://github.com/eddyem/stm32samples/tree/master/F1:F103/Canon_managing_device}}
(fig.~\ref{canbuscanon}). Device have simple letter-based text protocol over USB, allowing
changing focus (relative and absolute) and diaphragm size, also it can switch lens between manual
and external control.
We were not the only ones who reverse-engineered the lens protocol
\footnote{\url{https://illunis.com/camera-products/clc-usb/}}, however, the products of our
``competitors'' are very thick, which does not allow their device to be used with many CCD light
detectors due to the large depth of the chip position.

Another popular commercial device actively used in our equipment is a variety of USB-controlled
``relay modules''. The only inconvenience was the management of the relay data: no one expected
that the miracle of Chinese industry did not have management software ``out of the box''. The
problem was solved in just a few working
days\footnote{\url{https://github.com/eddyem/eddys_snippets/blob/master/USBrelay}}.
Similar relays are used among other things in the ``Astro-M'' complex to turn on lighting inside
a dome-space or artificial ``flat field'' (fig.~\ref{screenrelay}).

\begin{pict}
    \twopicts{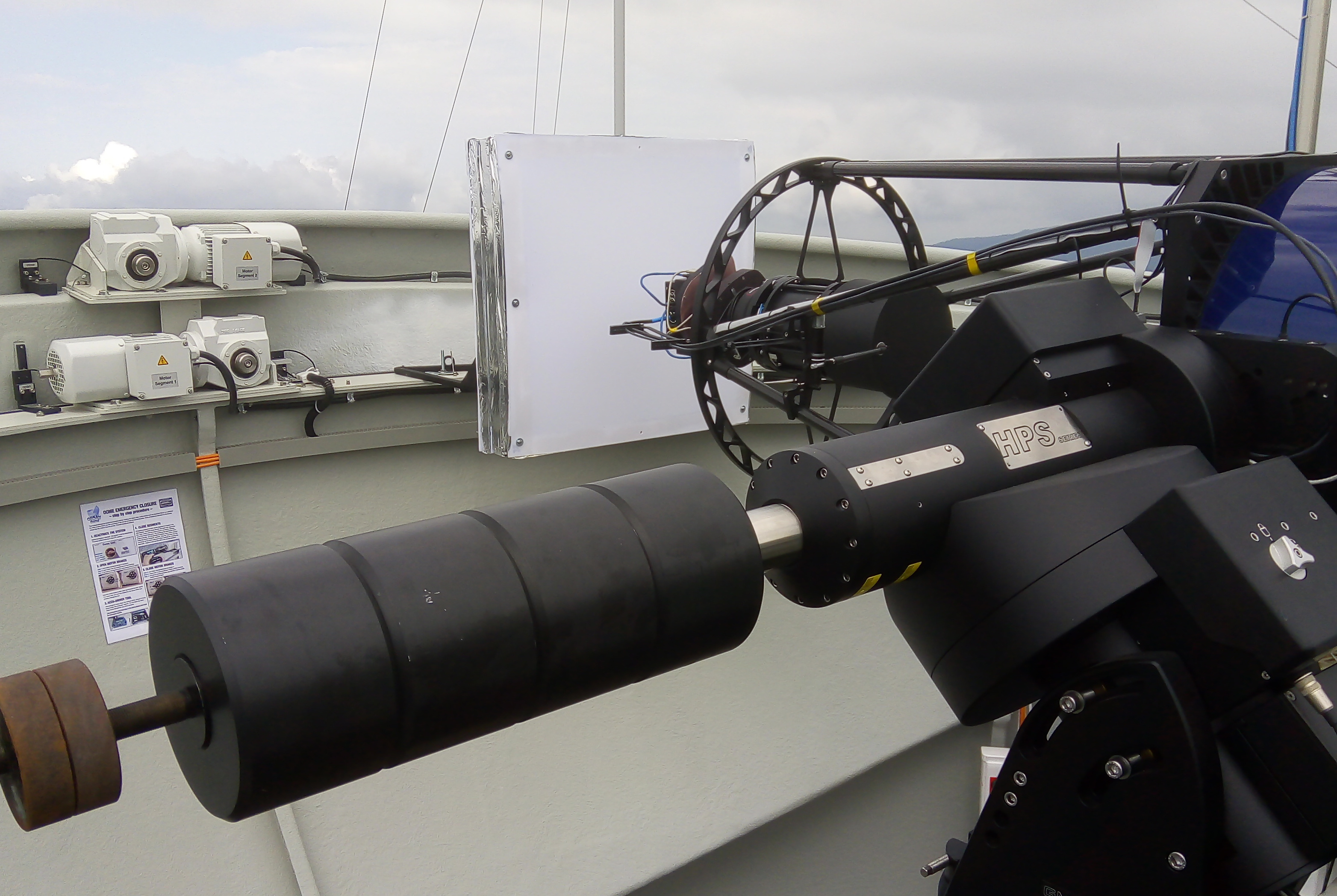}{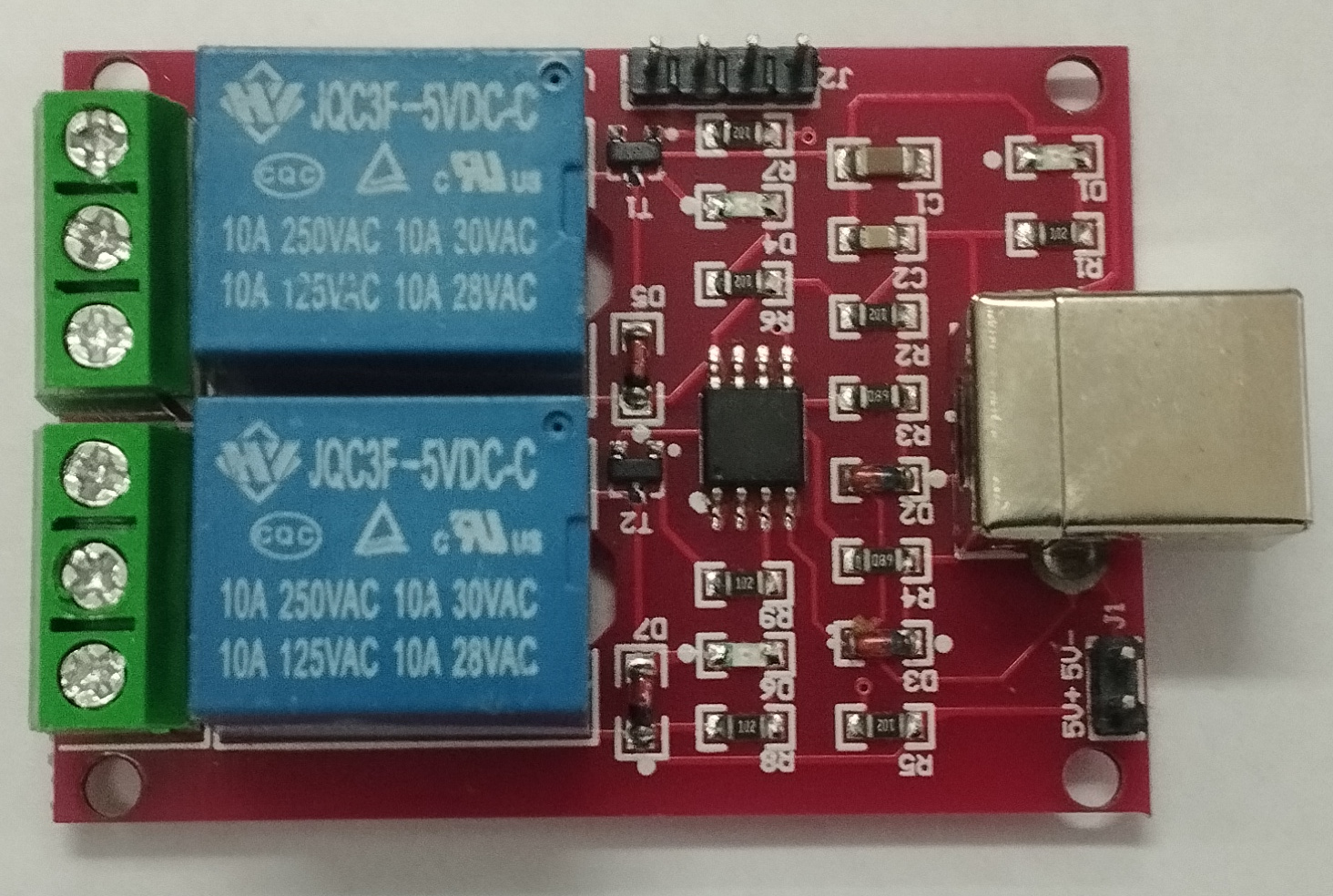}
    \caption{Telescope \No\,1 of ``Astro-M'' complex with flat-field screen (left), commercial USB
    relay module (right).}
    \label{screenrelay}
\end{pict}

However, it is not always possible to control the load via USB. Already at distances of more than
three to five meters the CAN-bus is more reliable. To control the calibration illuminance module
of the fibre-fed spectrograph, a special
unit\footnote{\url{https://github.com/eddyem/stm32samples/tree/master/F0:F030,F042,F072/usbcan_relay}}
 was developed (which also has manual control via push-buttons). The device allows to switch
 on\slash off a couple of relays, read data from two ADC inputs, manage three PWM outputs,
 switch on\slash off four LEDs. It have simplest letter-based text protocol over USB and
 ``non-standard'' CAN protocol (first byte is command, another are data).

One of our latest developments is an automatic liquid nitrogen supply
device\footnote{\url{https://github.com/eddyem/stm32samples/tree/master/F3:F303/NitrogenFlooding}}.
It is still in the prototype stage. The device has USB or CAN-bus control, as well as manual
control. For manual control, a compact screen and a set of buttons are installed. The level of
liquid nitrogen in the dewar is determined by the line of thermistors. The supply is carried out by
boiling nitrogen using a heating element. In case of overflow or at the end of the supply the air
valve is activated releasing the pressure in the dewar.

\section{Investigations of BTA and Zeiss-1000 with TeA}
TeA showed itself well in carrying out ``technical'' observations on BTA and Zeiss-1000. When
installing it on the BTA, it became necessary to supplement the device with a module with a
Foucault knife. Unfortunately, the current layout of the device does not allow automation of such
measurements; it is necessary, at a minimum, to remove the control computer from the device
flange to free up more space for additional equipment. But the development of this unit is planned
for the future, and will probably be carried out in 2026--2027

\begin{pict}
    \includegraphics[width=\textwidth]{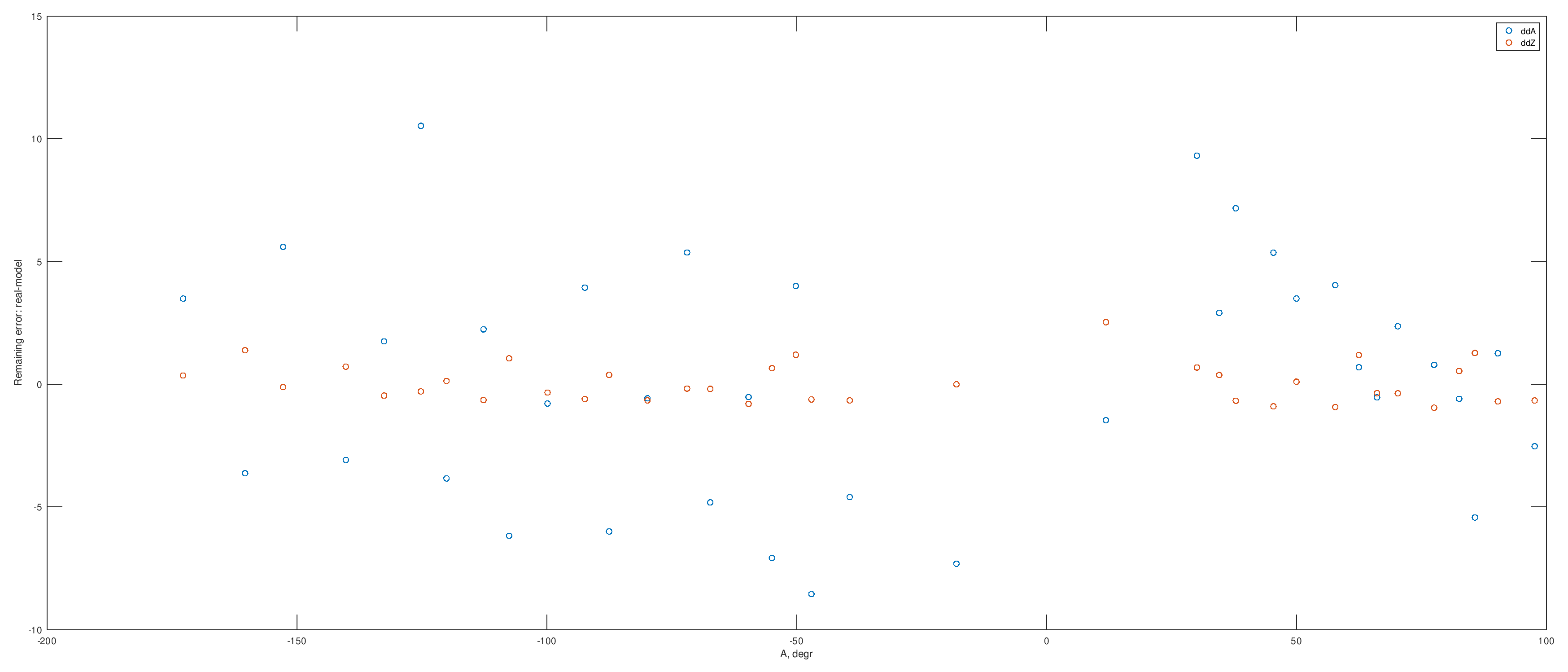}
    \caption{Residual errors after PCS calculations on unbalanced BTA. Blue~-- azimutal, red~--
        zenith errors.}
    \label{difA}
\end{pict}

\subsection{Pointing correction}
The mechanics of any telescope mount are imperfect. And the larger the telescope, the more
problems with pointing and tracking the object. When the telescope began to be controlled by
computer, they tried to solve this problem by introducing a pointing correction system (PCS) based
on natural factors. These factors depend on the type of telescope mount and characterize
inaccuracies in setting encoders' zero points, inaccuracy in adjustment and bending of axes, etc.
To determine the current PCS coefficients appropriate observations are carried out periodically
(usually once every half-year) during which pointing errors at test objects are calculated using as
many reference points as possible (preferably at least a hundred). Observations at the primary
focus are fully automated, but data collection at the Nasmyth foci still have to be done
manually  (perhaps in the future, this process will also be automated on the new N2~local
corrector).

Just 20 years ago, all such observations were performed literally by hand. It was necessary to aim
at each point manually entering its coordinates and then move the star using the telescope
correction remote control to the required centre (usually marked with a piece of plasticine on the
monitor screen). Nowadays, this procedure is fully automated and requires about 4.5\,hours
to obtain data from a hundred or so fields.
Pointing errors are calculated depending on the size of the light detector: on large fields
conventional astrometry is carried out using \texttt{solve-field} procedure from the package
\texttt{astrometry.net}, and on small fields it is measured the deviation of the brightest star (in
this case manual intervention is often required if automatic pointing by catalogue of $10^{m}$
stars, a brighter star  may appear in the field).

The huge problem of classical PCS is that many coefficients are linearly dependent. This
technique was suitable for ``pre-computer'' measurements, because the coefficients reflected the
real characteristics of mechanical systems that could be measured without resorting to
observations. Nowadays, there is an urgent need to reconsider these methods.

Classic PCS coefficients for an alt-azimuth mount are calculated as follows:
$$dA = K_0 + K_1/\tan Z + K_2/\sin Z - K_3\sin A/\tan Z + K_4\cos\delta\cos P/\sin Z,$$
$$dZ = K_5 + K_6\sin Z + K_7\cos Z + K_3\cos(A) + K_4\cos\phi\sin A,$$
where $Z$~is zenith angle, $A$~-- azimuth, $\delta$~-- declination, $P$~-- parallax angle,
$\phi$~is the latitude of observing place, $K_0$~-- azimuth zero, $K_1$~-- horizontal axis
inclination, $K_2$~-- collimation error, $K_3$~-- latitude error of vertical axis, $K_4$~-- time
error, $K_5$~-- zenith zero, $K_6$~-- sine component of tube bend, $K_7$~-- cosine component of
tube bend.

As we can see, finding coefficients based on observational results is an ill-posed problem. And
although in the case of a well-balanced telescope it converges completely by successive
approximations and least squares methods, in the general case it is unsolvable.

\begin{pict}
    \centerline{\hbox to 0.5\textwidth{\hss\includegraphics[width=0.6\textwidth]{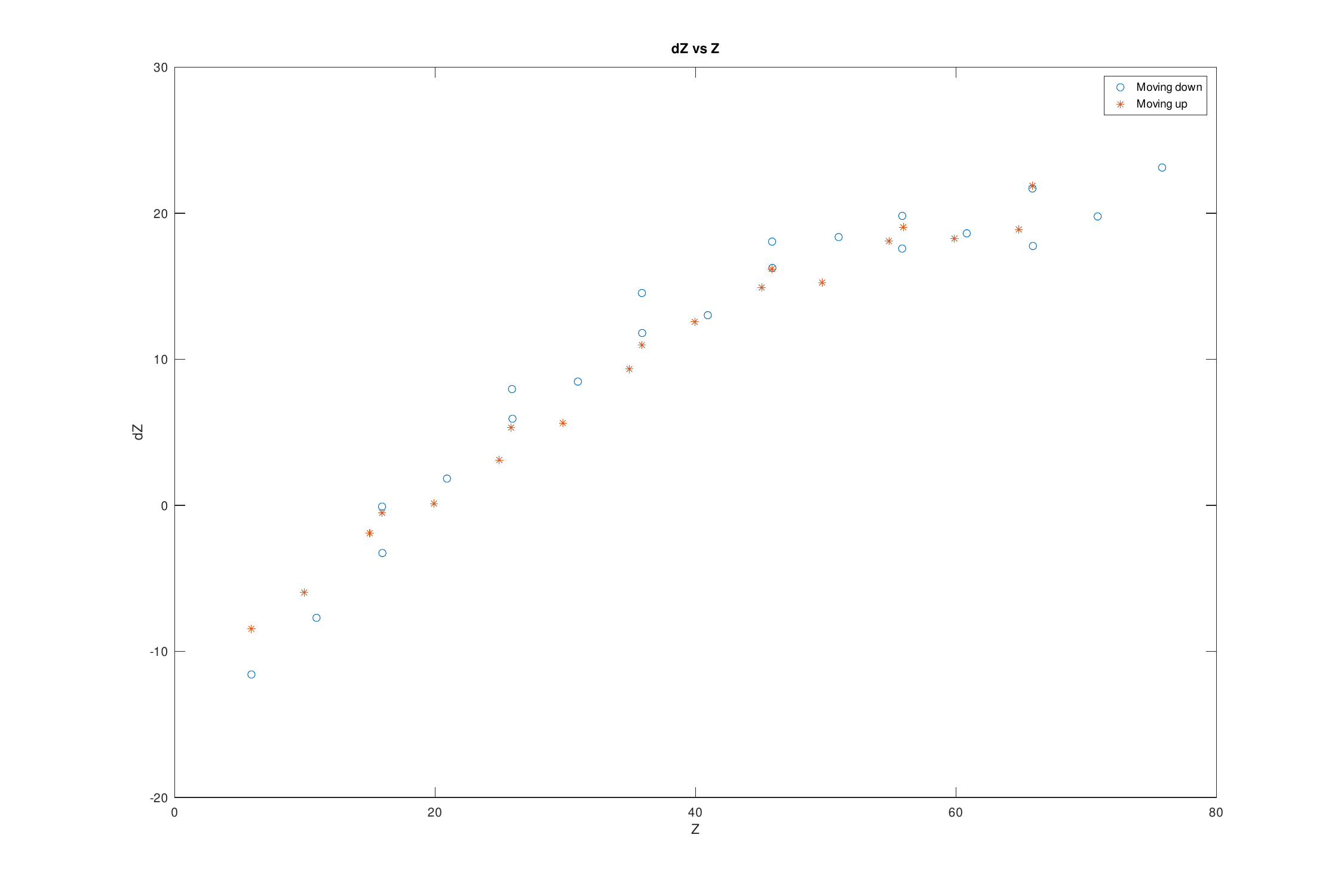}}\hfil
        \hbox to 0.48\textwidth{\hss\includegraphics[width=0.6\textwidth]{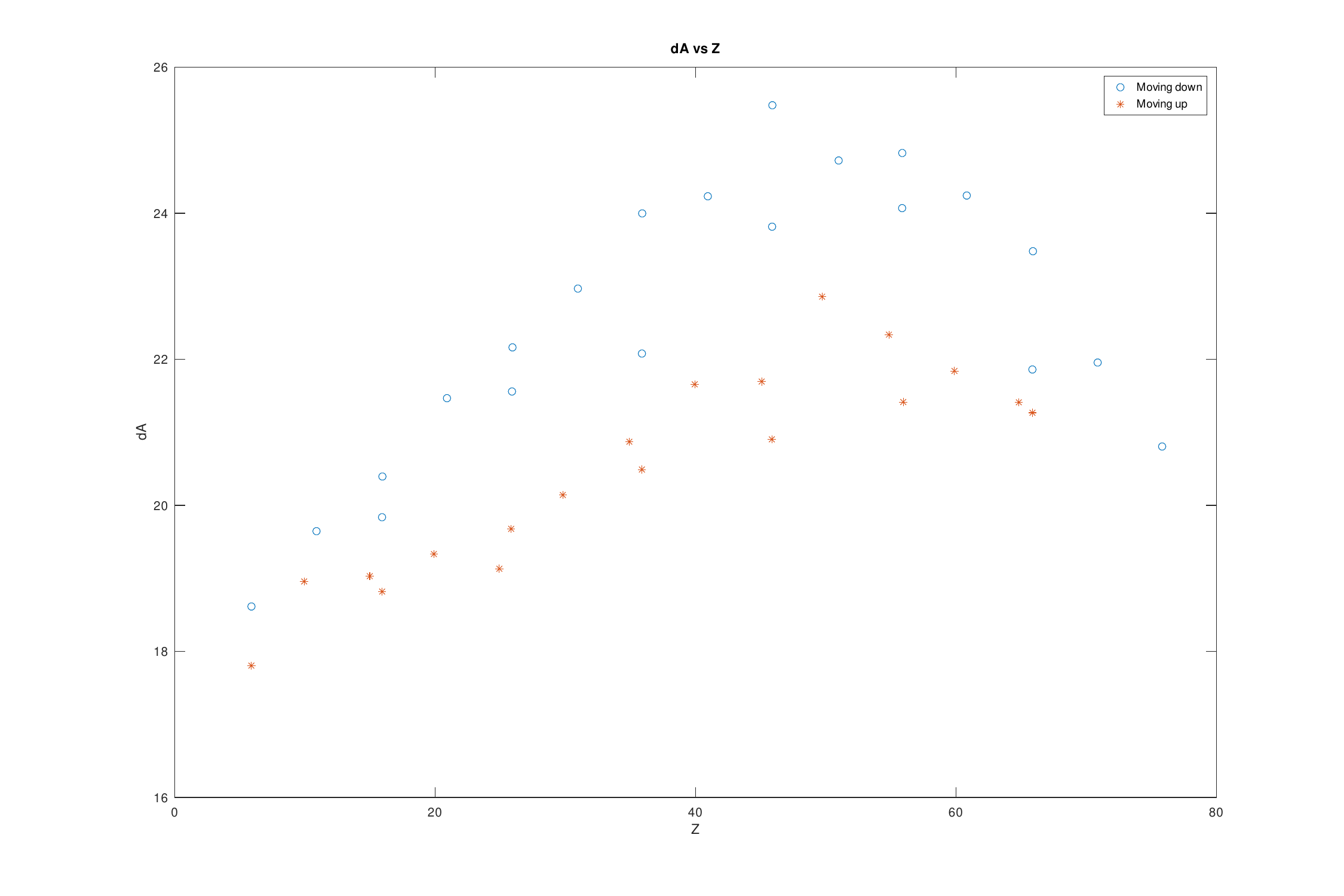}\hss}}
    \caption{BTA main mirror hysteresis measured by star positions (may, 2024); left: by zenith axis,
        right: by azimuthal axis. Blue
        circles~-- tube moving down, red asterisks~-- tube moving up.}
    \label{hyster}
\end{pict}

\begin{pict}
    \includegraphics[width=0.8\textwidth]{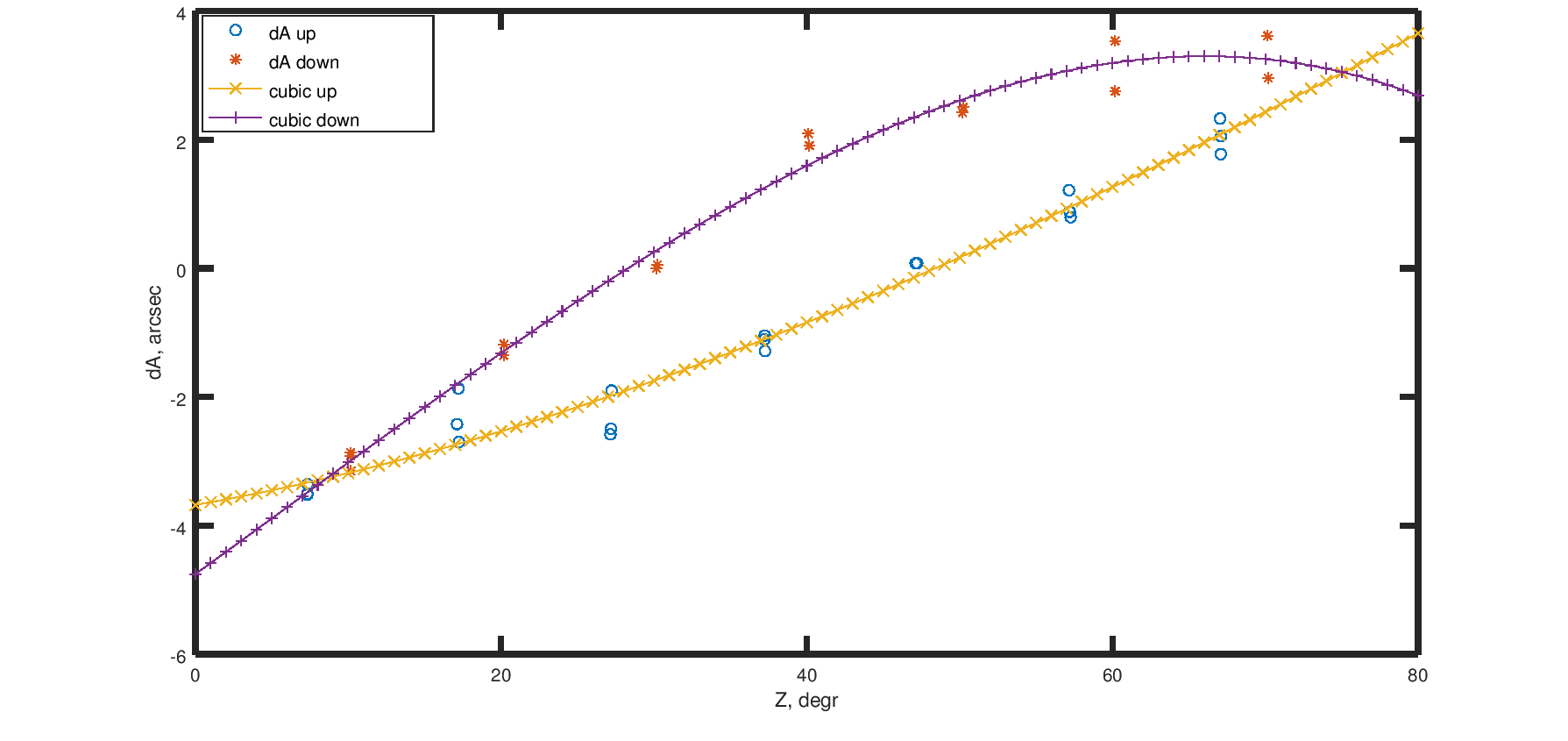}
    \caption{BTA  main mirror hysteresis measured by in-frame micrometers (august, 2017).}
    \label{hyster2}
\end{pict}

Let's take the following example. In May 2024, a comprehensive modernization of the BTA
N2~focus instruments began. After disassembling part of the equipment despite the installation of a
compensating weight the telescope turned out to be so unbalanced that it was even clearly visible
using a micrometer installed in the shank of the azimuthal axis. After carrying out measurements to
calculate the PCS it turned out that a clear sinusoid is observed in the residual errors which
cannot be eliminated using only classical coefficients (fig.~\ref{difA}).

In the literature, it is proposed to use specific orthonormal polynomials instead of classical
coefficients~\cite{PCS}. In the future, we hope to turn into a similar system of polynomials on all
telescopes: both alt-azimuth and equatorial.

\begin{pict}
    \twopicts{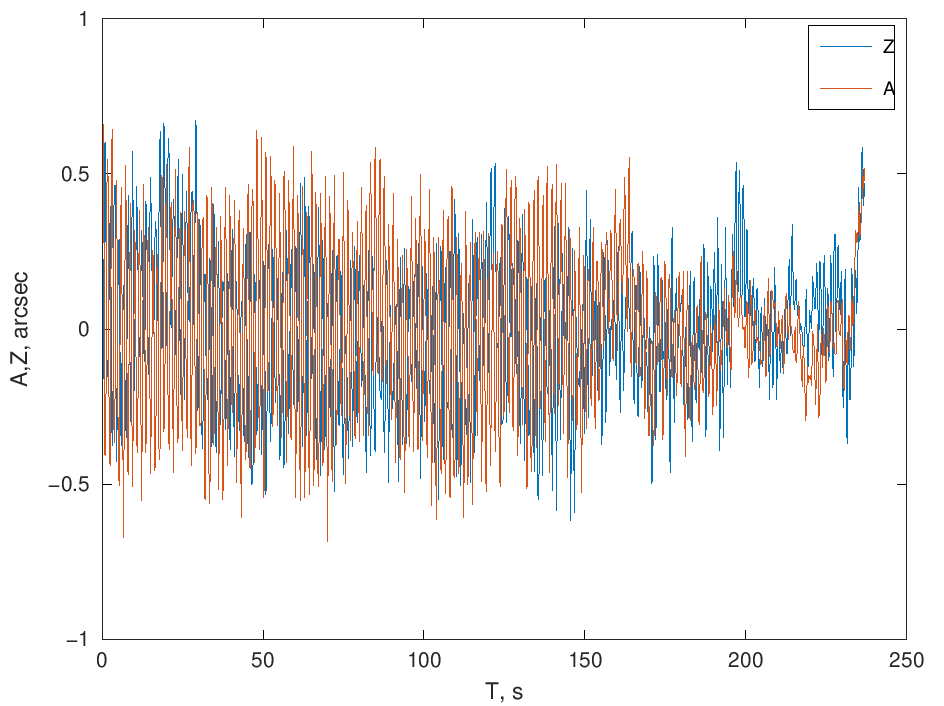}{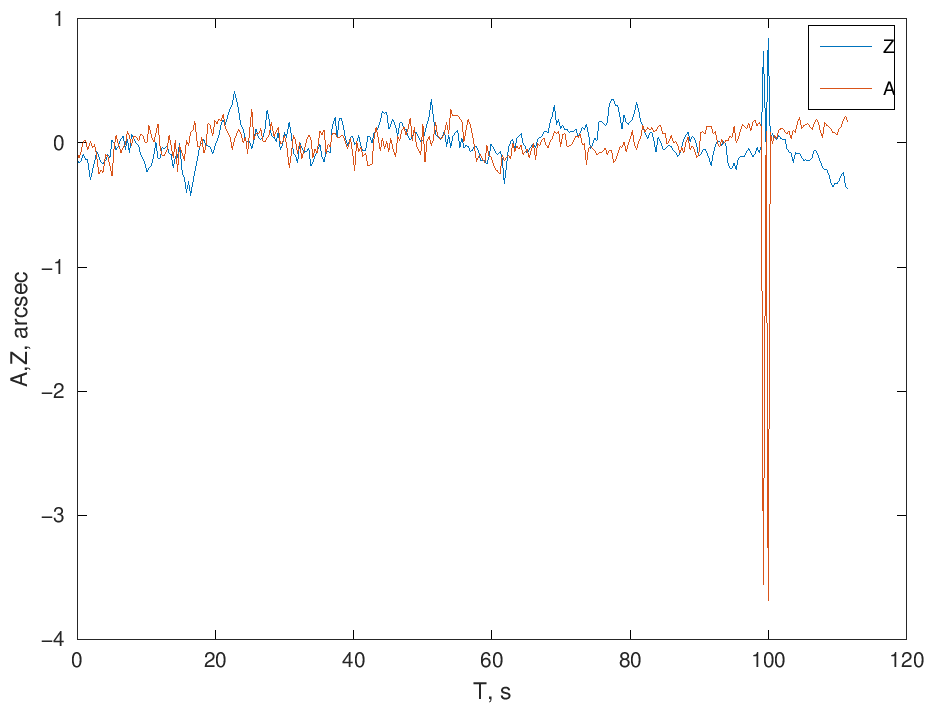}
    \caption{Vibration of star image in primary focus of BTA. Left: when guiding system is on, right:
    when telescope is fully off
    (pointing to Polar star).}
    \label{tubeflutter}
\end{pict}
\begin{figure}
    \includegraphics[width=0.9\textwidth]{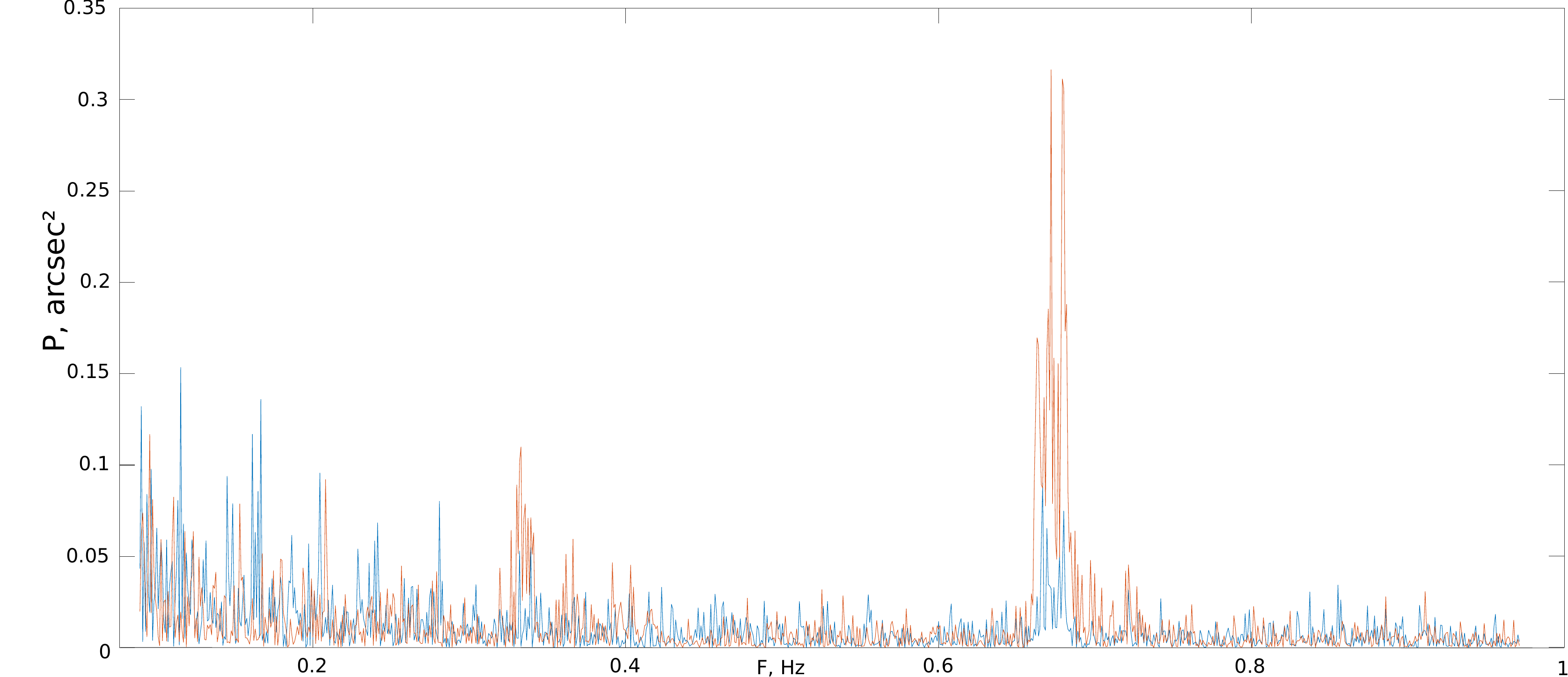}
    \caption{BTA tube vibrations power spectra (from left of fig.~\ref{tubeflutter}). Blue~--
    Z-component, red~-- A-component.}
    \label{freq}
\end{figure}

\subsection{Mirror hysteresis and mount natural frequencies}
Even PCS studies carried out on a well-balanced telescope showed that the residual deviations in
azimuth were not less than~$3''$. The reason for these deviations is the hysteresis of main
mirror (MM)  position in the frame. Measuring the tilt of MM using micrometers mounted on the
frame confirmed the conclusions drawn from observations across the sky. Judging by the results
the MM gravity center is shifted so that its main slopes occur relative to the vertical axis leading to
an azimuth shift of the image~(fig.~\ref{hyster}, \ref{hyster2}).

Another problem of such big mechanical construction like BTA is its natural frequencies.
Measurements showed the presence of natural frequencies of about 0.7\,Hz~(fig.~\ref{freq}).
Moreover, the oscillations are significantly increased in such positions of the telescope where
movement along the Z-axis has  the lowest speed (culminations, area of the north pole). Apparently,
the swinging of the tube in~Z  is also reflected in the azimuthal deviations of the image.

Measurements with the control system completely turned off reflect another effect on the images:
``gravitational waves''~--- density waves arising in the atmosphere leading to a displacement of
the image as a whole. Also right plot of fig.~\ref{tubeflutter} displays a strong image deviation
(over~$3''$ by Az) near $T=100\,$s. Most likely, it is caused by microseismics since there were no
gusts of wind during the observations.

The less massive telescope Zeiss-1000 did not detect any significant natural oscillations.

\begin{pict}
    \twopicts{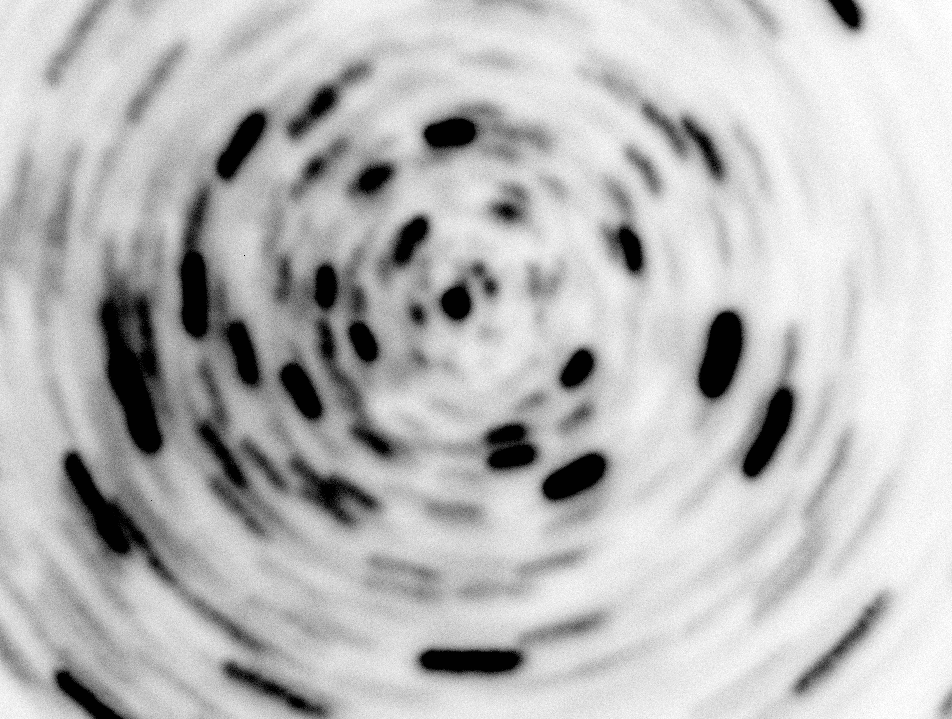}{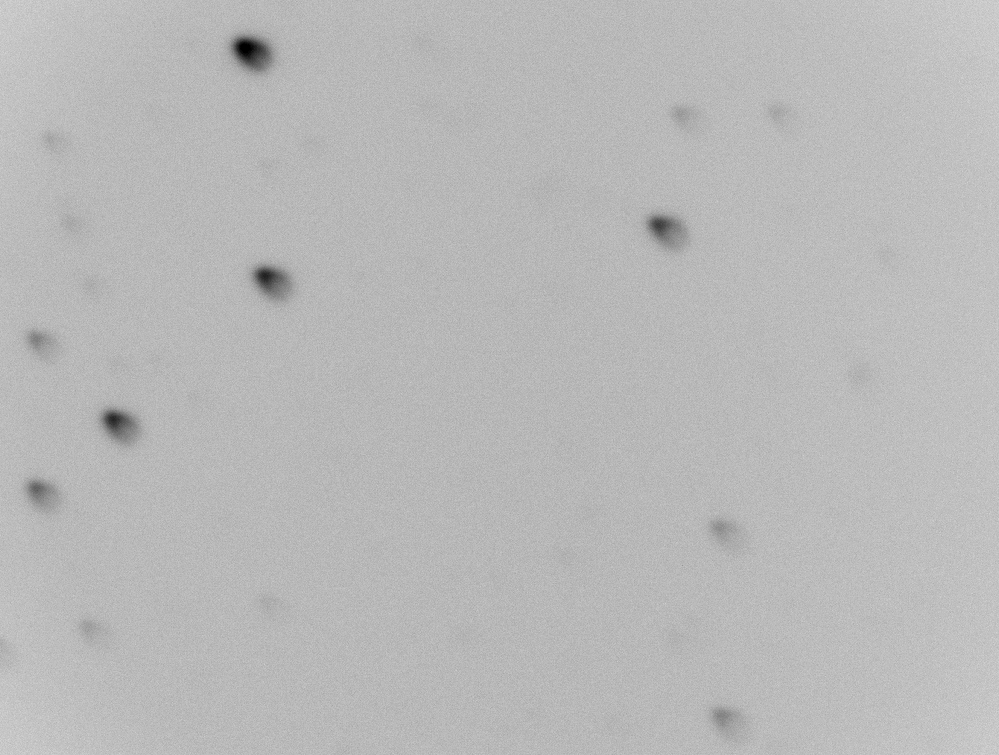}
    \caption{Long-exposured image during BTA primary focus flange rotation (left).
        One of frames with coma at the distance near 10\,cm from center of BTA field in primary focus
        (right).}
    \label{btacoma}
\end{pict}

\subsection{Aberrations and aberration axis position}
Part of the task was to determine the aberration centre of the BTA MM. Due to the large field of
view, at the greatest distances of the light receiver from the centre of the field the coma effect
is quite significant (fig.~\ref{btacoma}), so it is possible to quite accurately determine the aberration
centre position as the point of coma axes intersection. These measurements  showed that at the
BTA the aberration  centre lies in a circle with a radius of less than~$10''$ from the field rotation
centre.

On the Zeiss-1000 after changes made to reduce scattered light, the field of view decreased so
much that it was not possible to detect any measurable coma. Therefore, measurements of the
most optimal aberration-free focus offset were also unsuccessful.

In addition to image aberrations caused by the atmosphere and air inside the dome, there are
long-term aberrations associated with temperature deformations of the MM. These deformations
are especially noticeable on monolithic glass BTA MM.
In~2019, 80~TSYS-01 thermal sensors with an accuracy of~$\pm0.05^\circ C$ were installed on
back surface of BTA MM.  The developed data collection
system\footnote{\url{https://github.com/eddyem/tsys01}} makes it possible to predict temperature
deformations. Without forced air circulation inside the MM frame, a constant difference of up
to~$2^\circ C$ between the upper and lower parts of MM was observed, which made a fairly
significant contribution to the astigmatism (measurements were carried out using a
Shack--Hartmann wavefront sensor). After the ventilation system was restored, this temperature
distribution became symmetrical relative to the geometric centre, thus introducing only defocus
(fig.~\ref{mirtemp}). As mentioned above, a more detailed analysis of the influence of temperature
distribution on the mirror surface shape can be achieved by taking focograms after  TeA
upgrading with the addition of an automatic module with Foucault knife.

\begin{pict}
    \twopicts{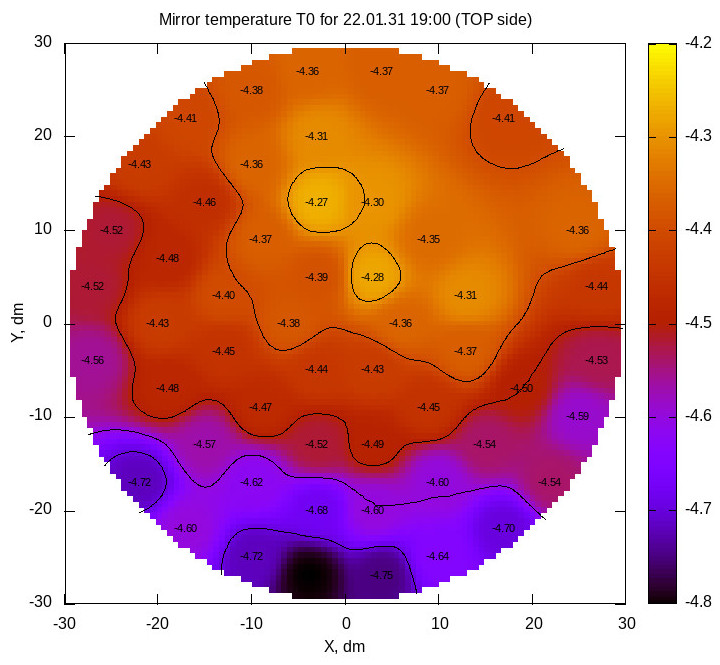}{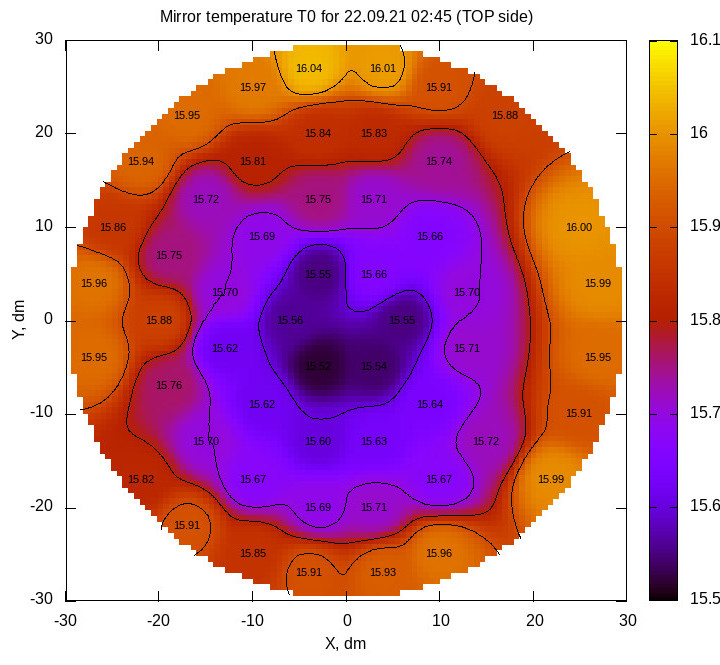}
    \caption{BTA main mirror temperature distribution: before (left) and after (right) active frame
        ventilation.}
    \label{mirtemp}
\end{pict}

\section{Towards BTA ACS modernisation}
The control system of BTA has not been modernized for more than 20~years. We are gradually
developing a comprehensive approach to its reorganization. It is planned to reduce the length of the
telescope's CAN-bus, replace outdated PEP controllers with compact monitoring
boards\footnote{\url{https://github.com/eddyem/stm32samples/tree/master/F3:F303/CANbus4BTA}}
for encoders and limit switches (fig.~\ref{ACS}), update drives and motors. Since modernization of
BTA must take place without stopping it, it will be done in stages. And at the first stage,
it is assumed that only the outdated PEP-controllers and some other units will be replaced with full
preservation of the topology and protocol. During further modernization, two control systems will
essentially work in parallel on the BTA, and the old one will be disabled only after the new one has
been fully debugged.

Modernization of relay circuits can also be carried out using ready-made units. So, the FX3U PLC
clones seemed quite tempting for the price. "Out of the box" these PLCs are completely useless,
because they cannot be programmed in a high-level language; the manufacturer offers only a poor
pseudo-graphical environment or a very limited text pseudo-programming language (thus, the
capabilities of MCU installed on the PLC are practically not realized at all). A minor
modification\footnote{\url{https://github.com/eddyem/stm32samples/tree/master/F1:F103/FX3U}}
allows them to be used as relay switching units in non-critical circuits (unfortunately, the lack of
interface protection does not allow their use everywhere). The modification consists of soldering on
the PCB a CAN bus level converter and completely changing the firmware. Modified firmware
allows to work with unit over RS-232 or CAN bus. Serial protocol is quite simple text setters\slash
getters (like in other our devices), CAN protocol also is our ``standard'' with 2-bytes command
code, one byte parameter, one byte error code and four bytes of data. The source codes freedom
allows anyone to implement on this ``PLC'' everything they need in a reasonable programming
language without any restrictions on the use of the MCU resources.

\begin{pict}
    \twopicts{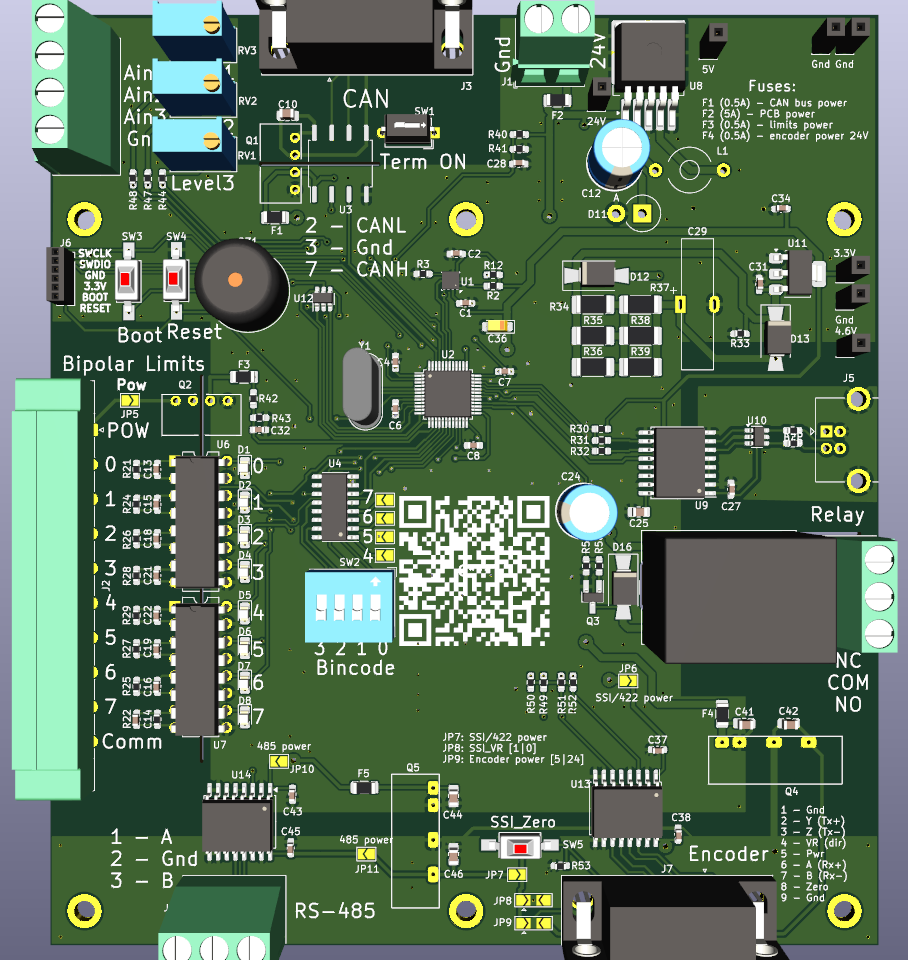}{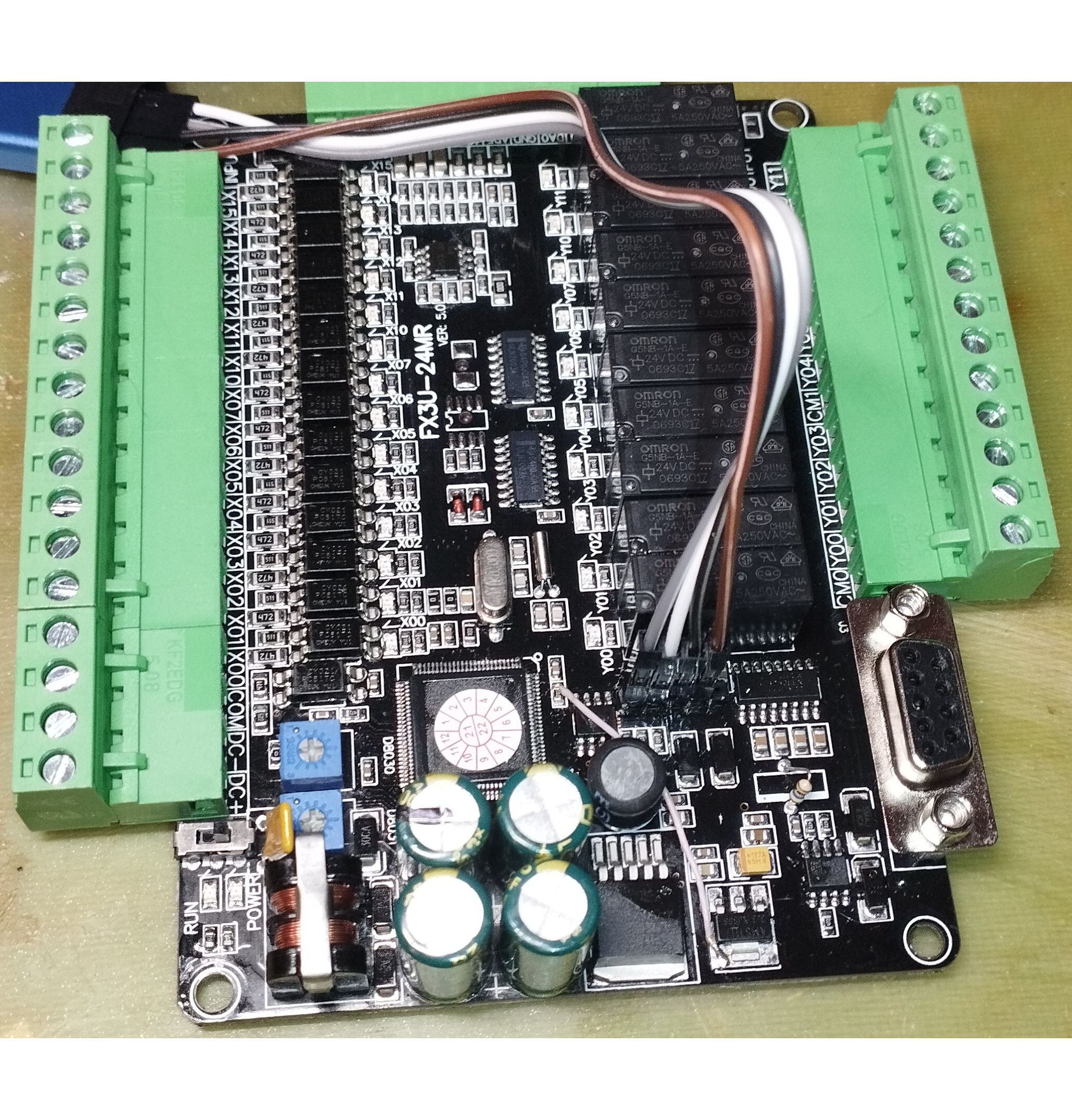}
    \caption{The developed ACS communication board (left), FX3U ``PLC'' clone (right).}
    \label{ACS}
\end{pict}

Alas, the policies of USA and its dependent countries towards Russia led to the impossibility
of legally purchasing many components. So we have to study Chinese analogues. For example,
some power frequency drives were purchased. Unfortunately, there are practically no CAN bus
drives on the white market; what is offered is mostly outdated and very limited in capabilities (for
example, the impossibility of multi-master on the bus) MODBUS. So, we have to develop
something to translate data between common ACS CAN bus and some MODBUS sectors.
The developed ACS module is equipped with a galvanically isolated RS-485, which will allow using
the unit to control equipment via MODBUS. In addition to the specified interfaces, the module can
be connected to a laptop via USB for debugging purposes. There is a possibility of polling three
analogue signals, eight limit switches and SSI or RS-422 rotary encoder. The installed relay allows
breaking the safety relay assembly if necessary. All external interfaces except the ADC are
equipped with galvanic isolation.

\section{``Astro-M'' 0.5-meter telescopes complex }
It is worth mentioning the ``Astro-M'': complex of five 50-cm telescopes, the construction of which
began in~2017. Nowadays, two telescopes of the complex are actively working and producing
scientific publications (e.g. \cite{exoplanets}, \cite{blazar} and so on). Our task is to put the
remaining three into operation, and in the future ensure their independent robot mode working.

All five telescopes were manufactured in Russia by the company ``Astrosib''.
The first two ``Astrosib'' telescopes are installed on a ``10-micron'' mount, which has all the
necessary functionality ``out of the box'' (fig.~\ref{astrom}). All that remained was to develop
auxiliary  daemons\footnote{\url{https://github.com/eddyem/small_tel}} working with equipment and
monitoring meteorological parameters. One of the difficulties was the development of a utility for
generating a list of pointing inaccuracies, from which the mount calculates PCS coefficients.
Mount documentation did not contain information about the algorithm for PCS calculating
(sign, epoch, etc.). Thanks to the help of 10-micron forum this problem was solved (the sources of
this utility you can also find in the repository).

\begin{pict}
    \twopicts{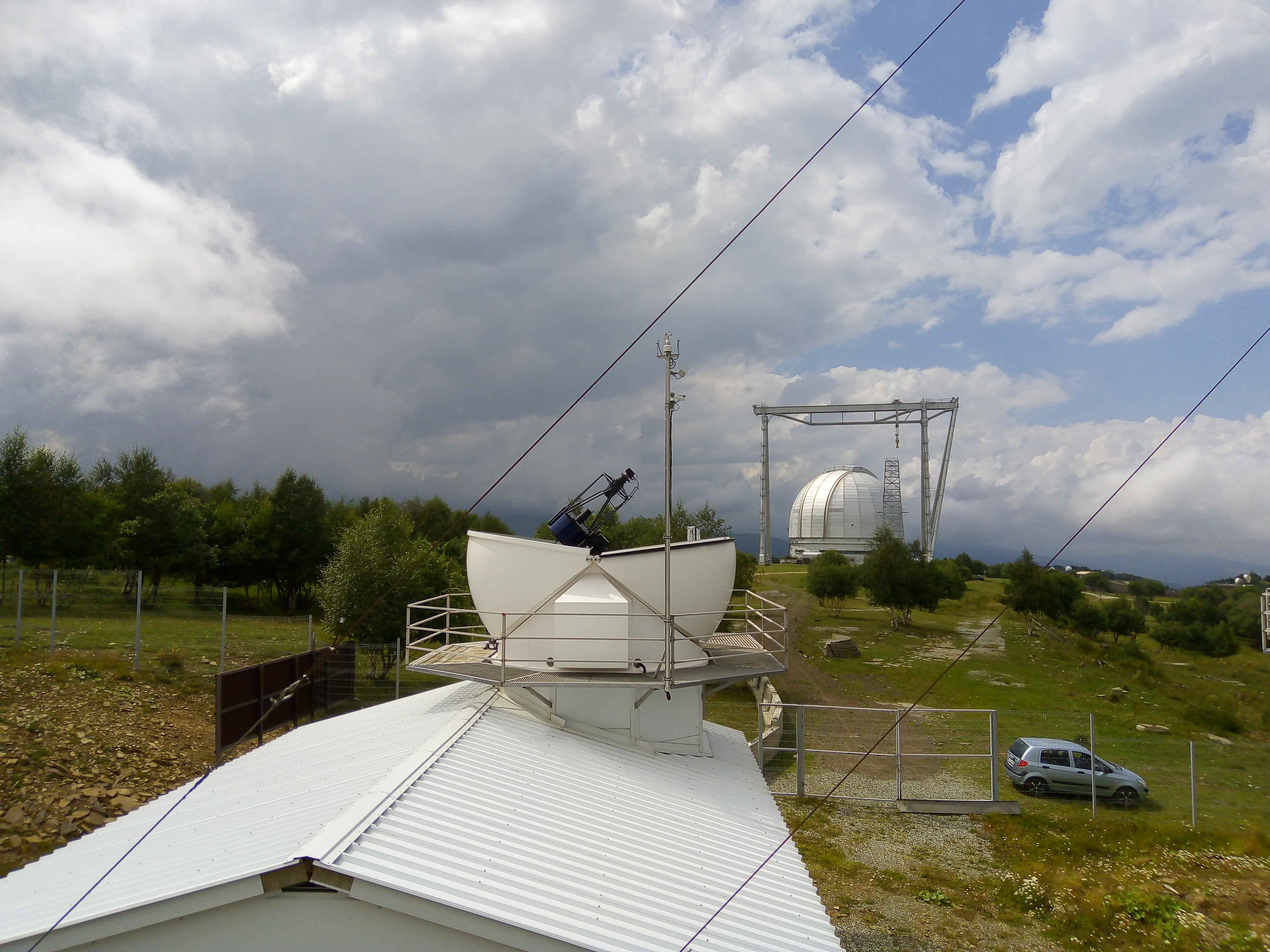}{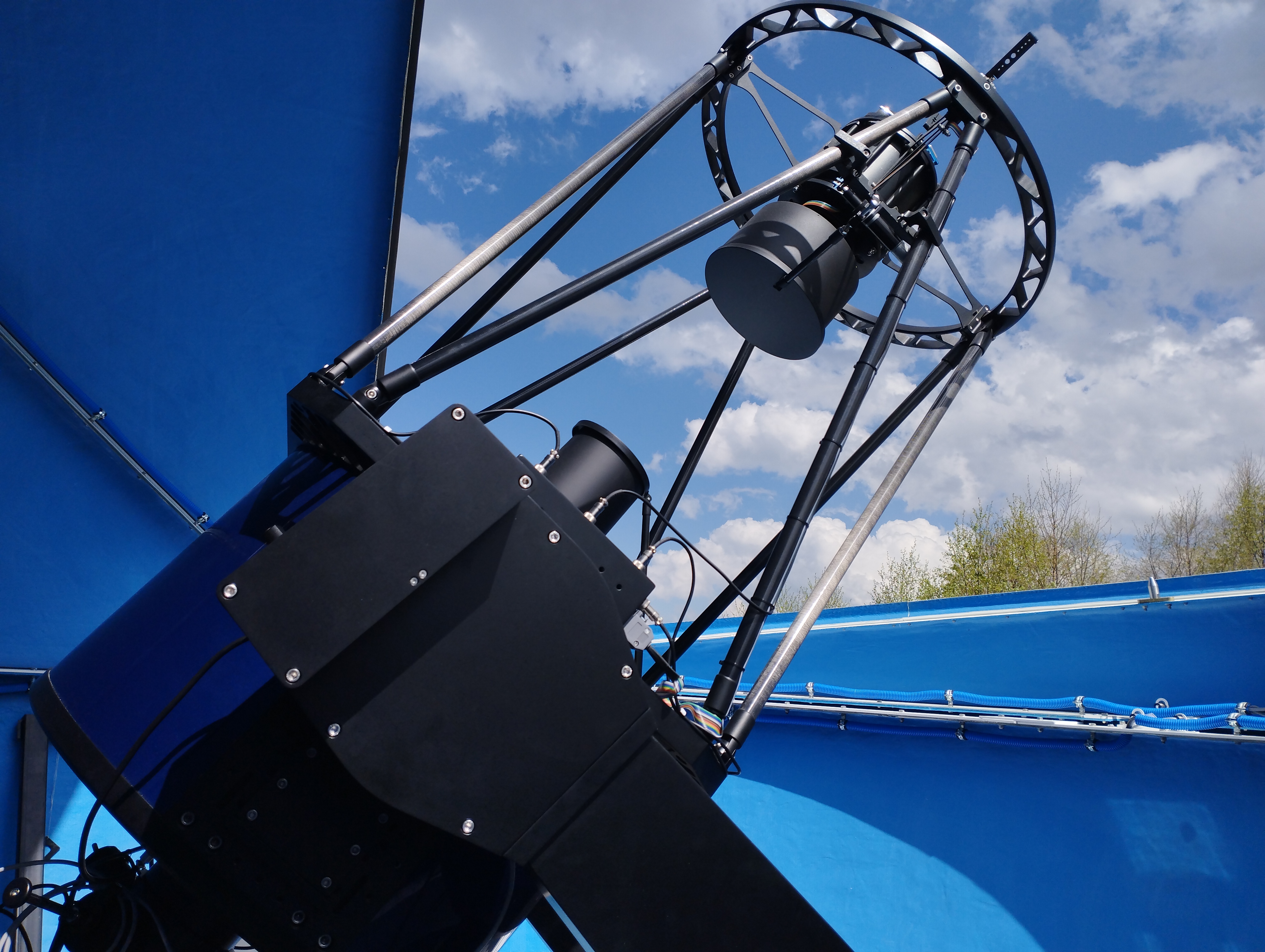}
    \caption{First telescope of ``Astro-M'' complex.}
    \label{astrom}
\end{pict}

All telescopes are installed in the all-sky domes: the first is ``Baader Planetarium'', the rest are
``Astrosib'' produced. Mounts, telescope nodes and domes are controlled via serial interface
RS-232. The main part of the protocols is described in the documentation, which allowed for a
relatively short period of time to implement the simplest network daemons that provide node
control. For security purposes, the daemons operate exclusively via UNIX sockets or local TCP
sockets. Authentication and communication with low-level daemons is the task of the scheduler,
common to all telescopes.

Both telescopes are equipped with simplest photometers based on FLI filter turrets. On the first one
there is a CCD FLI PL9000, on the second there is a Raptor EM-CCD.
When using the FLI CCDs, it turned out that they were not equipped with LEDs for illuminating the
chip, which made it impossible to get rid of ``spirits'' from previous exposures. ``Spirits'' from very
bright stars were sometimes observed even on the fifth to seventh frame.
Three CCDs purchased for small telescopes were upgraded by us, thus equipped with two
850\,nm and two 940\,nm LEDs (fig.~\ref{fliccd}). LED control is connected to one of the GPIO
outputs. So each exposition starts from 2-second dark frame with LEDs turned on following by
three bias frames with $32\times32$ binning and one dark frame with binning~$2\times2$ (all with
LEDs turned off). It has been experimentally established that this particular ``ritual'' allows one to
completely get rid of spirits even from very bright objects that cause the charge to spread
throughout the entire reading column.

\begin{pict}
    \includegraphics[width=0.5\textwidth]{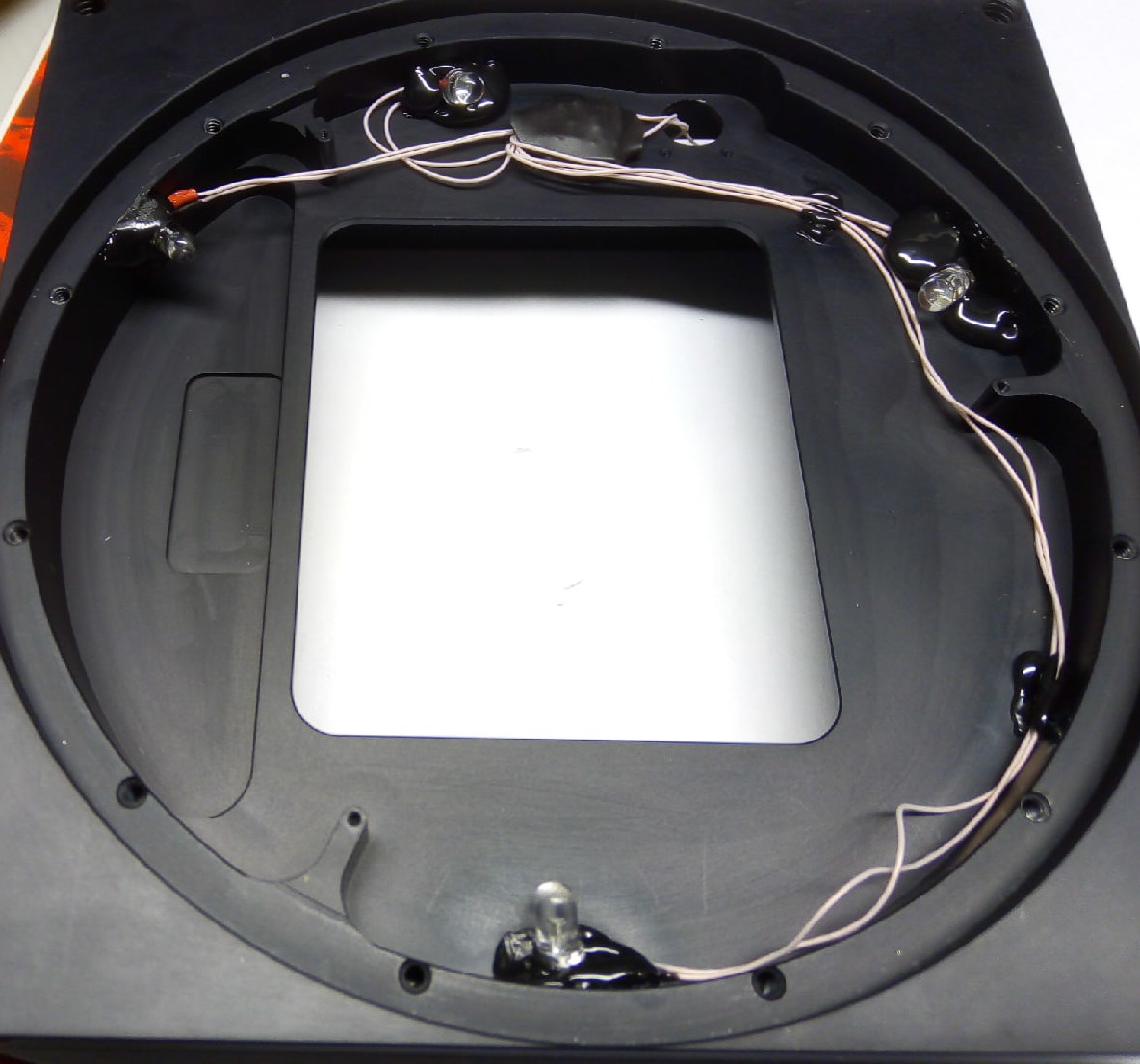}
    \caption{FlI CCD modernisation with four IR LEDs.}
    \label{fliccd}
\end{pict}

Due to the lack of an infrared all-sky camera, there is  no reliable way to fully robotize the
``Astro-M'' complex, so it still operates in semi-automatic mode. Every night the operator makes a
decision on the possibility of observations and, if it is positive, launches the main script.
While there is no general scheduler, we use elementary bash scripts for interaction between
daemons and utilities that control the equipment. The main script tracks the time until telescope
should be relocated, the time until sunrise and weather conditions. At the observations end the
script parks telescope, closes dome and launches reference exposures  (flat, bias and dark), after
which it turns off equipment power and produce archiving of gathered data (the size of archived
night is from 4\,GB in summer to 9\,GB in winter thus taking near 750--800\,GB per year). The
transition to a fully automatic mode of operation implies writing a full-fledged scheduler that will not
only be able to independently determine the possibility of observations start, but also in the case of
several objects per night select the most suitable ones based on the observation conditions.

The remaining three telescopes are mounted on an ``Astrosib'' fork mount equipped with ``Sidereal
Servo~II'' controllers. Unfortunately, there is no SDK for GNU/Linux, or even software for
configuring these controllers. In fact, we will have to implement functionality completely similar to
``10-microns''.

\end{document}